\documentclass[12pt]{article}
\usepackage{graphicx}
\usepackage{amssymb,epsfig}
\def\bseq{\begin{subequation}}  
\def\eseq{\end{subequation}}
\def\bsea{\begin{subeqnarray}}  
\def\esea{\end{subeqnarray}}

\newcommand{\bbox}{\lower.2ex\hbox{$\Box$}}

\newcommand{\beq}{\begin{equation}}
\newcommand{\eeq}{\end{equation}}
\newcommand{\bea}{\begin{eqnarray}}
\newcommand{\eea}{\end{eqnarray}}
\newcommand{\ena}{\end{eqnarray}}

\newcommand {\non}{\nonumber}

\renewcommand{\a}{\alpha}
\renewcommand{\b}{\beta}

\newcommand{\pa}{\partial}
\newcommand{\g}{\gamma}
\newcommand{\G}{\Gamma}

\newcommand{\e}{\epsilon}
\newcommand{\z}{\zeta}

\newcommand{\m}{\mu}

\newcommand{\p}{\pi}

\newcommand{\Db}{\bar{D}}

\newcommand{\Phib}{\bar{\Phi}}

\newcommand{\Del}{\nabla}

\newcommand{\hu}{|h_1|^2}
\newcommand{\hd}{|h_2|^2}

\newcommand{\htr}{|h_3|^2}

\begin{document}
\begin{titlepage}
\begin{flushright}
IFUM-- 904 --FT \\
Bicocca--FT--07--15\\

\end{flushright}

\vspace{.4cm}
\begin{center}
\noindent{\Large \bf
Conformal Invariance and Finiteness Theorems for}\\
\vspace{.2cm}
\noindent{\Large \bf Non--Planar $\beta$--deformed $\mathcal{N}=4$ SYM Theory}
\end{center}
 \vspace{0.4cm}

\vfill
\begin{center}

{\large \bf Federico Elmetti$^1$,~Andrea Mauri$^{1,2}$ and Marco Pirrone$^3$} \\

\vspace{0.5cm}

{\small $^1$ Dipartimento di Fisica, Universit\`a di Milano and\\
INFN, Sezione di Milano, Via Celoria 16, I-20133 Milano, Italy\\

\vspace{0.1cm}
$^2$ Department of Physics, University of Crete, 71003 Heraklion,
Greece\\

\vspace{0.1cm}
$^3$ Dipartimento di Fisica, Universit\`a di Milano--Bicocca and\\
INFN, Sezione di Milano--Bicocca, Piazza della
Scienza 3, I-20126 Milano, Italy}\\

\end{center}
\vfill
\begin{center}
{\bf Abstract}
\end{center}
We study the conformal invariance of non--planar $\beta$--deformed
$\mathcal{N}=4$ SYM theory using the coupling constant reduction (CCR)
formalism. We show that up to order $g^{10}$, differently from the
planar case, we can remove the scheme dependence in the definition of
the theory without reducing to the real $\beta$ case. We also compute
the gauge beta function up to four loops and see that the generalized
finiteness theorems proposed in [hep-th: 0705.1483] still hold.
\vspace{2mm} \vfill \hrule width 3.cm
\begin{flushleft}
e-mail: federico.elmetti@mi.infn.it\\
e-mail: andrea.mauri@mi.infn.it\\
e-mail: marco.pirrone@mib.infn.it
\end{flushleft}
\end{titlepage}

\section{Introduction}

Marginal deformations of $\mathcal{N}=4$ super Yang--Mills theory have
recently drawn much attention in the context of conformal
generalizations of AdS/CFT correspondence. The so--called
$\beta$--deformation is an interesting example of this class of
theories thanks to the work of Lunin and Maldacena \cite{LM} where its
gravity dual description has been found. From the field theory point
of view this deformation is realized by enlarging the space of
parameters of the original $\mathcal{N}=4$ theory with the following
modification of the superpotential:
\beq
i\,g \,\mbox{Tr}\left(\Phi_1
\Phi_2 \Phi_3 - \Phi_1 \Phi_3 \Phi_2\right)\,\,\longrightarrow
\,\,i\,h\,\mbox{Tr}\left(e^{i\pi \beta}\,\Phi_1 \Phi_2 \Phi_3 -
e^{-i\pi \beta}\,\Phi_1 \Phi_3 \Phi_2\right) \eeq

\noindent
where $h$ and $\beta$ are two new complex coupling constants in
addition to the gauge coupling $g$, which is chosen to be real. The
resulting theory preserves $\mathcal{N}=1$ supersymmetry and it is
expected to become conformally invariant only if a precise relation
among the coupling constants exists \cite{LS}. Several papers have
been devoted to the study of an explicit realization of this condition
in the planar case (\cite{MPSZ}-\cite{StiKaz}). Keeping $\beta$ real,
the Leigh--Strassler constraint turns out to be satisfied at all order
in perturbation theory by the exact solution $h\bar{h} = g^2$
\cite{MPSZ}. The case of complex $\beta$ requires a more careful
investigation since the conformal condition gets perturbatively
corrected. In order to properly describe the fixed point surface in
the space of couplings, the coupling constant reduction (CCR) program
has shown to be a powerful tool (\cite{oehme}-\cite{JZ}). Using this
approach, in \cite{Noi} it is claimed that conformal invariance and
scheme independence of the theory can not be achieved at the same time
for the complex $\beta$ deformed case in the planar limit\footnote{The
possible scheme dependence of the vanishing $\g$ condition has been
first noted by the authors of \cite{RSS2}. In \cite{Noi} we explicitly
considered this feature and studied its implications.}.\\ The aim of
this paper is to achieve a better understanding of the problem by
looking at the finite $N$ case (see also \cite{FG}-\cite{Voi}). Working
perturbatively we will ask for the chiral and gauge beta functions to
vanish in order to define the theory at its conformal point. In
Section 2 we will analyze the properties of the two--point chiral
correlator. Once again we will make use of the CCR procedure to obtain
the vanishing of the anomalous dimension. This amounts to express the
chiral couplings as functions of the gauge coupling $g$. As a
consequence the perturbation theory is naturally defined in terms of
powers of $g$ instead of powers of loops. At order $g^6$ we meet the
first non--trivial situation because at this stage different loop
diagrams start contributing at the same order in $g$. We will see that
up to order $g^{10}$, differently from the planar case, there is
enough freedom to remove the scheme dependence without reducing to the
real $\beta$ case.

Then we will turn to consider the gauge beta function. As CCR approach
allows different loop orders to mix, it is not obvious that standard
finiteness theorems \cite{PW,GMZ} should hold. So, having canceled the
chiral beta function up to $\mathcal{O}(g^7)$ does not automatically
imply the vanishing of the gauge beta function at
$\mathcal{O}(g^9)$. The fact that this is still the case is a highly
non--trivial check that we will cover in details in Section 3. The
same problem was studied in \cite{Noi} in the planar case where it was
shown by an explicit computation that the condition for the vanishing
of the anomalous dimension $\g$ at $\mathcal{O}(g^8)$ actually ensures
the vanishing of the gauge beta function at
$\mathcal{O}(g^{11})$. This result was obtained making use of
background field method combined with covariant
$\nabla$--algebra. However it is worth noting that the procedure
followed in \cite{Noi} is not the standard one (extensively explained
in \cite{GZ}), which turned out to be too involved. Here, working at a
lower order in $g$ but keeping $N$ finite, we will be able to get
through the calculation adopting both of the methods and
checking explicitly the equivalence of the two.

\section{Chiral Beta Function and Conformal Condition}

Let us consider the ${\cal N}=1$ $\b$--deformed action written in
terms of the superfield strength $W_\a= i\Db^2(e^{-gV}D_\a e^{gV})$, where $V$ is a real prepotential,
and three chiral superfields $\Phi_i$ with $i=1,2,3$, all in the adjoint representation of the $SU(N)$ gauge group. With notations
as in \cite{superspace} we have

\bea
S &=&\int d^8z~ {\rm
Tr}\left(e^{-gV} \Phib_i e^{gV} \Phi^i\right)+ \frac{1}{2g^2}
\int d^6z~ {\rm Tr} (W^\a W_\a)\nonumber\\
&&+ih  \int d^6z~ {\rm Tr}( ~q ~\Phi_1 \Phi_2 \Phi_3 - \frac{1}{q}~
\Phi_1 \Phi_3 \Phi_2 ~)
\nonumber \\
&& + i\bar{h}\int d^6\bar{z}~ {\rm Tr} ( ~\frac{1}{\bar{q}}~ \Phib_1
\Phib_2 \Phib_3 - \bar{q} ~\Phib_1 \Phib_3 \Phib_2~ )\qquad\qquad
q\equiv e^{i\pi\b}  \label{actionYM}
\eea
\\
Here $h$ and $\b$ are
complex couplings and $g$ is the real gauge coupling constant.
In the undeformed  ${\cal N}=4$ SYM theory one has $h=g$ and $q=1$.
From now on we will be considering 't Hooft rescaled quantities

\beq\label{resc}
h\,\rightarrow\,\frac{h}{\sqrt{N}}\qquad
g\,\rightarrow\,\frac{g}{\sqrt{N}}
\eeq
\\
in order to easily make contact with the planar limit.
Moreover we notice that the phase of $h$ can always be reabsorbed by a
field redefinition, so that the effective  number of independent real parameters in the
superpotential is actually three. For later convenience we choose them to be $\hu,\,\hd$ and $\htr$, where

\beq \label{h123}
h_1 \equiv h \,q \qquad \qquad h_2 \equiv \frac{h}{q} \qquad
\qquad h_3\equiv h \,q-\frac{h}{q}
 \eeq \\
In the spirit of \cite{LS}
the idea is to find a surface of renormalization group fixed points in
the space of the coupling constants. To this end one can consider the
coupling constant reduction program (\cite{oehme}-\cite{JZ}) and express the
renormalized Yukawa couplings in terms of the gauge one:
\newpage
\bea
\label{expansion} && \hu=a_1 g^2+a_2 g^4+a_3 g^6+\dots \nonumber\\
&& \hd=b_1 g^2+b_2 g^4+b_3 g^6+\dots \\ && \htr=c_1 g^2+c_2 g^4+c_3
g^6+\dots \non \eea \\
This operation has an immediate consequence: we are forced to work
perturbatively in powers of $g$ instead of powers of loops.  To single
out a conformal theory we will ask for the chiral and gauge beta
functions to vanish.  In this section we will concentrate on $\b_h$
and adopt dimensional regularization within minimal subtraction
scheme. The chiral beta function is proportional to the anomalous
dimension $\g$ of the elementary fields and the condition $\b_h=0$ can
be conveniently traded with $\g=0$.  Even working in a generic scheme,
one can easily convince oneself that at a given order in $g^2$ the
proportionality relation between $\b_h$ and $\g$ gets affected only by
terms proportional to lower order contributions to $\g$. Therefore, if
we set $\g=0$ order by order in the coupling, we are guaranteed that
$\b_h$ vanishes as well \cite{JJN1}.  So the object we will be mainly
interested in is the two--point chiral correlator.

In \cite{Noi} this issue has been analyzed by considering the planar
limit where only two independent real constants enter the color
factors, namely $\hu$ and $\hd$.  As a result the definition of the
conformal theory was found to be scheme dependent as long as $\b$ was
complex.  In the non--planar case all of the three parameters enter
the calculation of the two--point chiral correlator.  We will see that
this difference will be important in the definition of the fixed point
surface.

The idea is to proceed perturbatively in superspace.  Supergraphs will
be evaluated performing the $D$--algebra inside the loops and the
corresponding divergent integrals will be computed using dimensional
regularization in $n=4-2\e$.  In this framework one could allow the
coefficients $a_i, b_i, c_i$ in (\ref{expansion}) to be expanded in
power series of $\e$ \cite{StiKaz}. Doing this, evanescent terms are
introduced {\it ad hoc} in order to deal with the $1/\e$ poles and
ensure the complete finiteness of the theory. However, after sending $\e
\rightarrow 0$, they do not enter the relation between renormalized
coupling constants so we will neglect their possible presence
hereafter.

Let us start at order $g^2$. As first proposed in \cite{RSS} it is
 convenient to consider the difference between divergent diagrams in
 the $\b$--deformed and in the ${\cal N}=4$ theory. This amounts to the
 evaluation of the chiral bubbles in Fig.1 which give the following
 divergent contribution to the chiral propagator

\beq\label{bob}
\frac{1}{(4\p)^2}~\left[
\hu+\hd-\frac{2}{N^2}\htr-2 g^2\right]~ \frac{1}{\e}~\left(\frac{\mu^2}{p^2}\right)^\e
\eeq
\\
where we have explicitly indicated
the factors coming from dimensionally regulated integral (here $p$ is the external momentum and $\m$ is the standard
renormalization mass).

\begin{figure} [ht]\label{one}
\begin{center}
\hspace{-1.3cm}\epsfysize=3,9cm\epsfbox{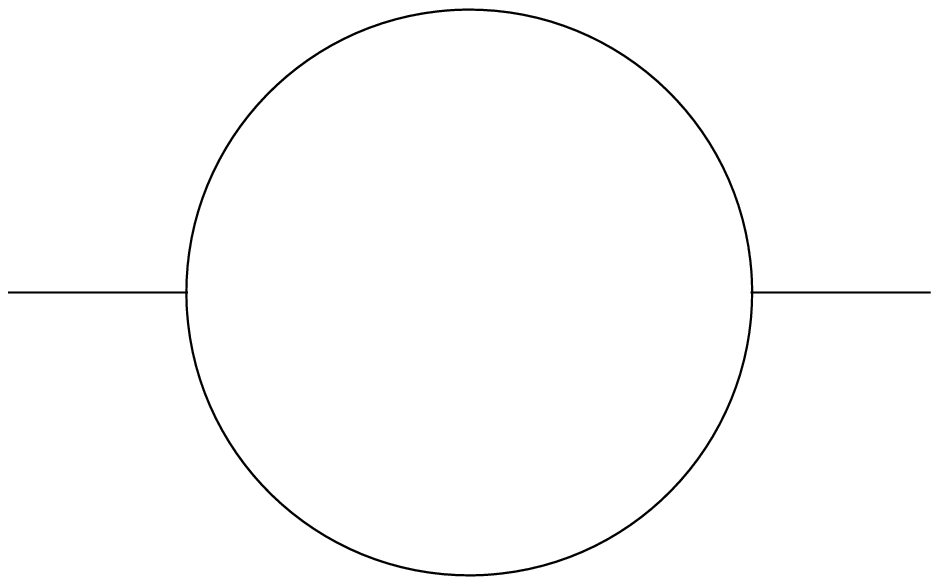}\hspace{0.5cm}\epsfysize=3,9cm\epsfbox{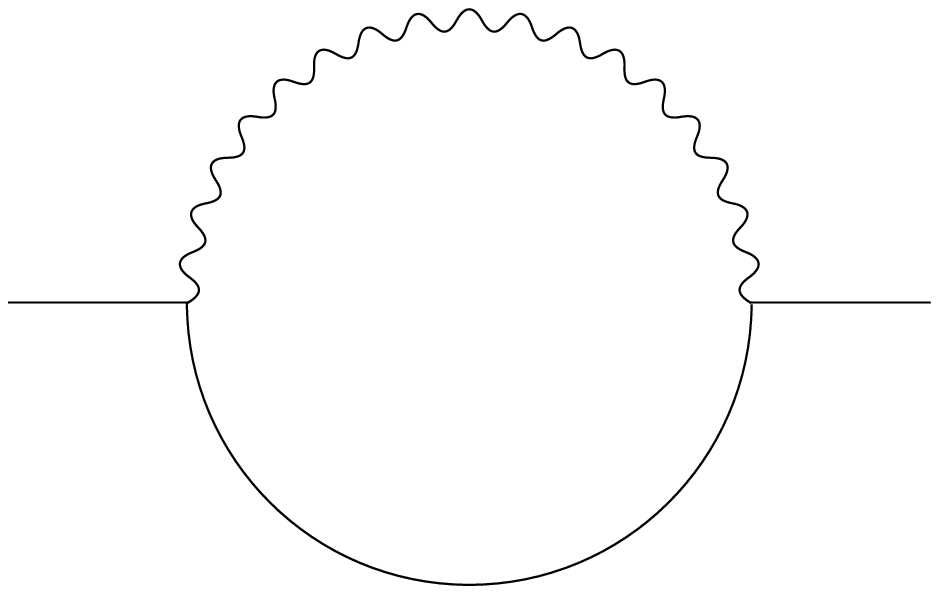}
\caption{One loop diagrams}
\end{center}
\end{figure}

\noindent
At this stage, in order to obtain a vanishing chiral beta function, the following condition has to be imposed

\beq
{\cal O}(g^2): \qquad \qquad  a_1+b_1-\frac{2}{N^2}c_1=2
\label{order1}
\eeq
\\
Moreover, it is well known that
\beq
\hu+\hd-\frac{2}{N^2}\htr=2 g^2
\label{finite}
\eeq
\\
ensures $\g=0$ up to two loops \cite{PSZ}.  So, looking at the chiral two--point contribution (\ref{bob}) at order $g^4$, we have the following additional requirement
\bea\label{g4}
{\cal O}(g^4): \qquad \qquad a_2+b_2-\frac{2}{N^2}c_2=0
\eea
\\
It is easy to see that equations (\ref{order1}) and (\ref{g4}) reduce
to the ones found in \cite{Noi} in the large $N$ limit.  When we move up
to the next order the situation becomes more involved with respect to
the planar case. In fact, working with finite $N$ we need to consider
the non--planar graph in Fig.2, whose contribution is:
\beq\label{cond3}
\frac{1}{(4\p)^6}~2\zeta(3)~\mathcal{F}~\frac{1}{\e}~\left(\frac{\mu^2}{p^2}\right)^{3\e}
\eeq
\\
where $\mathcal{F}\equiv\mathcal{F}\left(\hu,\hd,\htr, N^2 \right)$ reads \cite{RSS,Voi}
\beq \label{effe}
\mathcal{F}=\frac{N^2-4}{N^4}~\htr~\left[\frac{N^2+5}{N^2}~|h_3|^4-3\htr(\hu+\hd)+3(\hu-\hd)^2\right]
\eeq

\begin{figure}[ht] \label{cro}
\begin{center}
\hspace{-1.3cm}\epsfysize=3,9cm\epsfbox{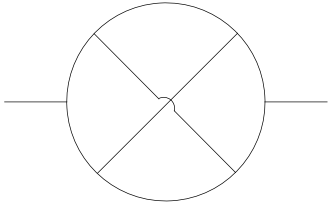}
\caption{Three loop non--planar diagram}
\end{center}
\end{figure}

\noindent
Notice that the color factor in (\ref{effe}) is suppressed as $1/N^2$ for large $N$.
Due to the expansion in (\ref{expansion}) both the one loop (\ref{bob}) and three loops (\ref{cond3}) structures
contribute  to the evaluation of $\g$ at  $\mathcal{O}(g^6)$.
The final result can be recast as

\beq
\frac{1}{\e}\left[ A\left( \frac{\m^2}{p^2}\right)^{\e} + \frac{B}{N^2}\left( \frac{\m^2}{p^2}\right)^{3\e}\right]
\label{forgamma}
\eeq
\\
where we have defined for concision

\beq\label{A}
A \equiv \frac{1}{(4\pi)^2}~(a_3 + b_3 - \frac{2}{N^2} c_3)
\eeq
\beq\label{B}
B \equiv \frac{2 \z(3)}{(4\p)^6}~ \frac{N^2-4}{N^2}~ c_1~\left[ \frac{N^2+5}{N^2}~c_1^2-3 c_1(a_1 + b_1)+3(a_1-b_1)^2\right]
\eeq
\\
The vanishing condition of the anomalous dimension at order $g^6$ can be read directly from the
finite $log$ term in (\ref{forgamma}):

\beq
{\cal O}(g^6): \qquad\qquad \qquad A+\frac{3 B}{N^2}=0
\label{gamma6}
\eeq \\
We emphasize that at this order the condition for the
vanishing of $\g$ and $\b_h$ is completely scheme
independent. However, from now on we will have to care about the scheme
dependence in the definition of the fixed points. To see this,
let us consider the counterterm needed at
this stage to properly renormalize the propagator in an arbitrary
scheme: \beq g^6 ~(A+\frac{B}{N^2}) ~(\frac{1}{\e} +\rho)
\label{counterterm}
\eeq
\\
where $\rho$ is a constant related to the choice of  finite renormalization. In fact,
if we were to push the conformal invariance condition one order higher we should compute the
chiral beta function at order $g^{9}$.  We expect to have several sources of nontrivial contributions
to $\g$ at this order:
one coming from the one--loop bubble proportional to $(a_4 + b_4-\frac{2}{N^2} c_4)$, then from
two--loop, three--loop and four--loop diagrams.  All of the diagrams containing subdivergences, namely
the two and four loop contributions, will be subtracted making use of the appropriate counterterms.
To be specific, a term like
\beq
g^{8} ~(A+\frac{B}{N^2}) ~(\frac{1}{\e} +\rho)~\frac{1}{\e} ~\left( \frac{\m^2}{p^2}\right)^{\e}
\label{counter2}
\eeq
\\
will appear in the calculation of $\g$. Therefore the request for vanishing anomalous dimension depends
unavoidably on the arbitrary constant  $\rho$ which appears in the form

\beq
(A+\frac{B}{N^2}) ~\rho
\label{scheme}
\eeq

\noindent
If we wanted to kill the scheme dependence of the result we would also need to impose the vanishing
of the combination $A+B/N^2$ which together with (\ref{gamma6}) would lead immediately to $A=B=0$.
The crucial observation is that in the non--planar case we deal with three parameters and at this stage
we have enough freedom to eliminate the scheme dependence from the conformal condition without reducing
to the real $\beta$ case.
In fact, the constraint $A=0$ gives

\beq
a_3+b_3-\frac{2}{N^2}c_3=0
\eeq
\\
while the condition $B=0$ combined with equation (\ref{order1}) yields\\

\beq
 \left\{
\begin{array}{lll}
a_1 & + & b_1=2 \non\\
c_1 & = &  0\non
\end{array}
\right.
\label{sol1}
\eeq\\
or, if $c_1 \neq 0$ \\
\beq
\left\{
\begin{array}[c]{lll}
a_1 + b_1  &  =  &   2\left(1 + \frac{c_1}{N^2}\right)
\\ \\
a_1- b_1  &  = &   \pm \, \sqrt{2\, c_1\,\left(1-\frac{N^2-1}{6\,N^2} \,c_1\right)}
\end{array}
\right.
\label{sol2}
\eeq
\\ These solutions allow for a non vanishing imaginary part of
$\beta$ (which is proportional to the combination $\hu-\hd$). At the
same time, they define the surface of renormalization fixed points
without any ambiguity related to the choice of regularization
scheme. It is clear that in the planar limit only the condition coming
from $A=0$ survives as the $B=0$ condition is subleading. So we are
left with $a_3+b_3=0$, in complete agreement with the result found in
\cite{Noi}.

If we move to the next order, a new scenario will show up. Having
imposed (\ref{sol1}) or (\ref{sol2}) only three graphs will contribute
to the anomalous dimension at order $g^8$ (Fig.3). Since these
diagrams are primitively divergent (no subdivergences are present) the
condition for $\g=0$ at this order turns out to be completely scheme
independent.
\begin{figure} [ht]
\begin{center}
\hspace{-1.3cm}\epsfysize=3,9cm\epsfbox{Bolla.eps}\epsfysize=3,9cm\epsfbox{Croce.eps}\hspace{1,4cm}\epsfysize=2.7cm\epsfbox{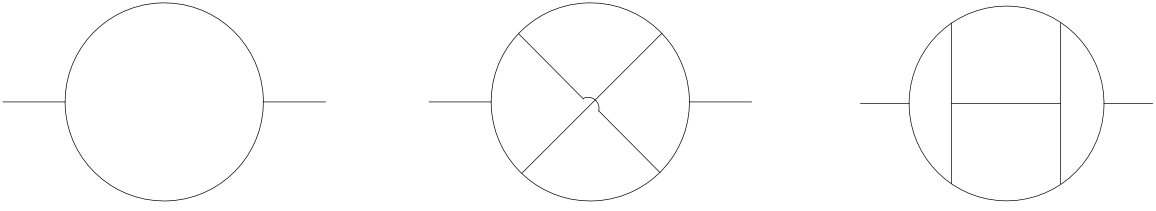}
\caption{Diagrams contributing to $\gamma$ at order $g^8$}
\end{center}
\end{figure}
In fact we have to consider the following expression:

\beq
\frac{1}{\e}\left[ A^\prime\left( \frac{\m^2}{p^2}\right)^{\e} + \frac{B^\prime}{N^2}\left( \frac{\m^2}{p^2}\right)^{3\e} + H \left( \frac{\m^2}{p^2}\right)^{4\e} \right]
\label{faivgamma}
\eeq
where we have denoted

\beq
A^\prime \equiv \frac{1}{(4\pi)^2} \left(a_4 + b_4 - \frac{2}{N^2}\, c_4\right)
\eeq
\beq
B^\prime \equiv \frac{6 \,\zeta(3)}{(4\pi)^6}\frac{N^2-4}{N^2}\left[
 (a_1-b_1)^2 c_2 + c_1 \Big(\frac{N^2-1}{N^2}\, c_1 c_2 + 4\,(a_1-b_1)(a_2-\frac{c_2}{N^2})-4c_2 \Big) \right]
\eeq
\beq \label{uxi}
 H \equiv -\frac{5\,\zeta(5)}{2(4\pi)^8}\,\left[ (a_1 - b_1)^4 + (a_1 + b_1)^4+
\frac{1}{N^2} \,f \Big(a_1,b_1,c_1,\frac{1}{N^2}\Big)-\frac{16(N^2+12)}{N^2}\right]
\eeq \\
\vspace{0,8cm}
\noindent
where $f$ can be read from Appendix A and we have used the relations (\ref{order1}) and (\ref{g4}). \hspace{1cm}
The vanishing of $\g$ reads
\beq\label{eq1}
\mathcal{O}(g^8):\qquad
\qquad \qquad A'+ \frac{3 B'}{N^2} + 4 H =0
\eeq
Again, in order to
remove scheme dependence from the $\mathcal{O}(g^{10})$ conformal
condition we have to impose:
\beq\label{eq2}
A'+ \frac{B'}{N^2} + H = 0
\eeq
At this stage, independently of the choice (\ref{sol1}) or
(\ref{sol2}), we have enough parameters to solve both equations
without restricting to the real $\beta$ case as in the planar
theory. On the other hand, if one sends $N \rightarrow \infty$, equations (\ref{eq1})
and (\ref{eq2}) reduce to the ones found in \cite{Noi}.
This large $N$ limit turns out to be smooth and does not present
any sort of singularity, so there is no contradiction between our results and
those found in \cite{Noi}.
We observe that a scheme--independent definition of the complex $\beta$
conformal theory can be achieved only thanks to subleading
coefficients which are projected out by the planar limit.

\vskip -2cm

\section{Gauge Beta Function and Finiteness Theorems}

Now we turn to consider the gauge beta function. Standard finiteness
theorems \cite{PW,GMZ} ensure the vanishing of $\beta_g$ at L+1--loops
once $\beta_h$ has been set to zero at L--loops. Here, as a consequence
of coupling constant reduction, we are forced to work order by order
in $g^2$ instead of loop by loop and it is not obvious that such
theorems still hold. Nevertheless in \cite{Noi} it was shown that in
the planar $\beta$--deformed theory the vanishing condition for $\beta_h$
at $\mathcal{O}(g^9)$ was sufficient to have vanishing $\beta_g$ at
$\mathcal{O}(g^{11})$.  This result was a strong indication that
finiteness theorems could be generalized as follows: if the matter
chiral beta function vanishes up to order $g^{2n+1}$ then the gauge
beta function vanishes as well up to order $g^{2n+3}$. Here we are
going to check this result at finite $N$ and for $n=3$. In order to do
this, we take advantage of covariant supergraph techniques combined
with background field method \cite{GZ}. The standard procedure
consists in looking at vacuum diagrams at a given perturbative order
and performing covariant $\Del$--algebra. Then by expanding propagators
one extracts tadpole type contributions with vector connections as
external legs.  Moreover one only selects diagrams containing at least
a $1/\epsilon^2$ pole (see \cite{GMZ} for details).  In the present
case, contributions to the gauge beta function at $\mathcal{O}(g^9)$
come from two and four loop vacuum diagrams (Fig.4).\\
\begin{figure} [t]
\begin{center}
\epsfysize=3,3cm\epsfbox{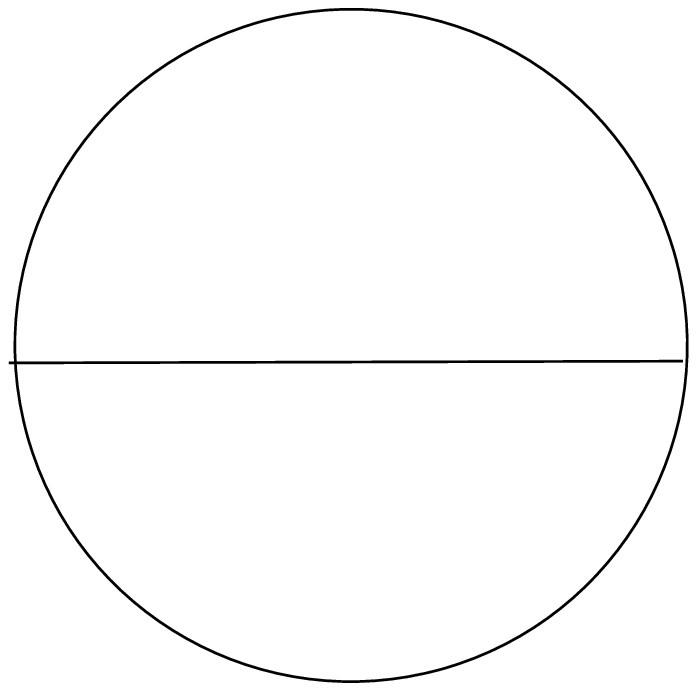}\hspace{3cm} \epsfysize=3,3cm\epsfbox{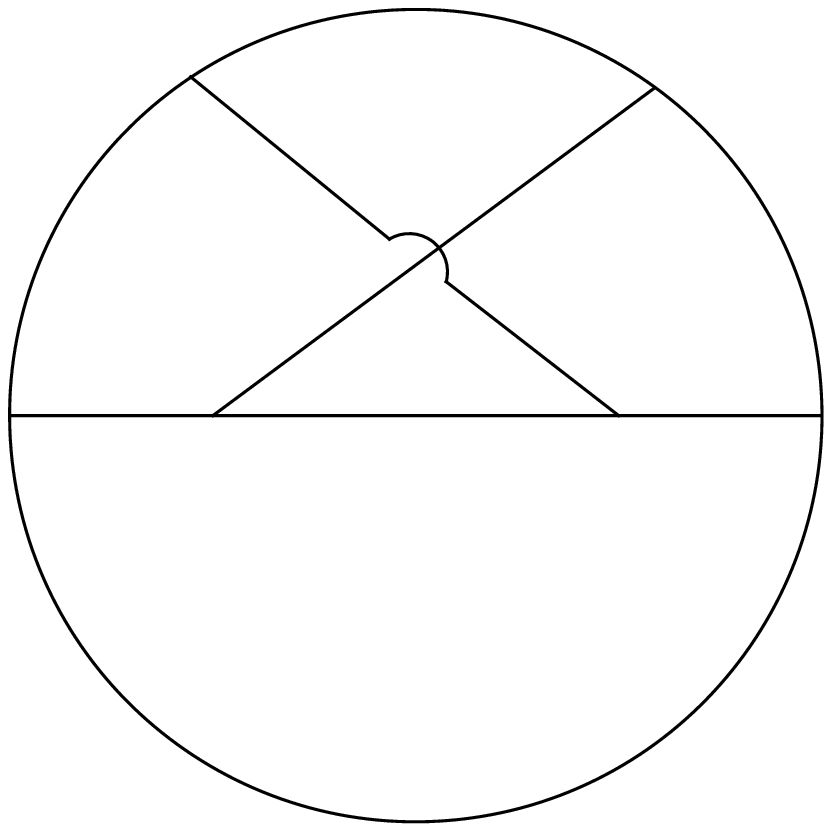}
\caption{Two and four loop vacuum diagrams}
\end{center}
\label{VACUUM}
\end{figure}
The analysis of the two loop diagram is straightforward and completely analogous to the one in \cite{GMZ}.
 Expanding the covariant propagators one obtains three times the
diagram in Fig.5 which corresponds to the term
 \beq
 \frac{1}{2}~{\rm{Tr} }~(\G^a\G_a) \int \frac{d^n k ~d^n q}{(2\p)^{2n}} \frac{1}{q^2(q+k)^2 k^4}
 \label{2loopdiv}
 \eeq
where $\Gamma_a$ is the vector connection.
\begin{figure} [h]
\begin{center}
\epsfysize=3,7cm\epsfbox{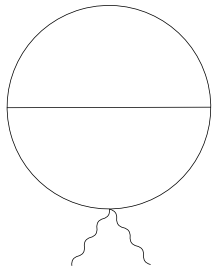}
\caption{Two loops tadpole diagram}
\end{center}
\label{gallo}
\end{figure}\\
 This integral contains a one--loop ultraviolet subdivergence and it is infrared
divergent. It is convenient to remove the
IR divergence using the $R^*$ subtraction procedure of \cite{CT}. After UV and IR subtractions one
isolates the  $1/\e^2$ term and rewrites the result in a covariant form,  obtaining the
following contribution to the two loop effective action:

\beq \label{2loop}
\frac{1}{(4 \pi)^2}\,\frac{3\,(N^2-1)}{4\,N}\, A\,
\frac{1}{\epsilon}\,\textrm{Tr}\int d^4 x\, d^2 \theta \,W^\alpha W_\alpha
\eeq\\
where we have inserted the A factor defined in (\ref{A}).\\
\begin{figure} [h]
\begin{center}
\begin{tabular}{ccc}
& \hspace{-0,8cm} \epsfysize=4,7cm\epsfbox{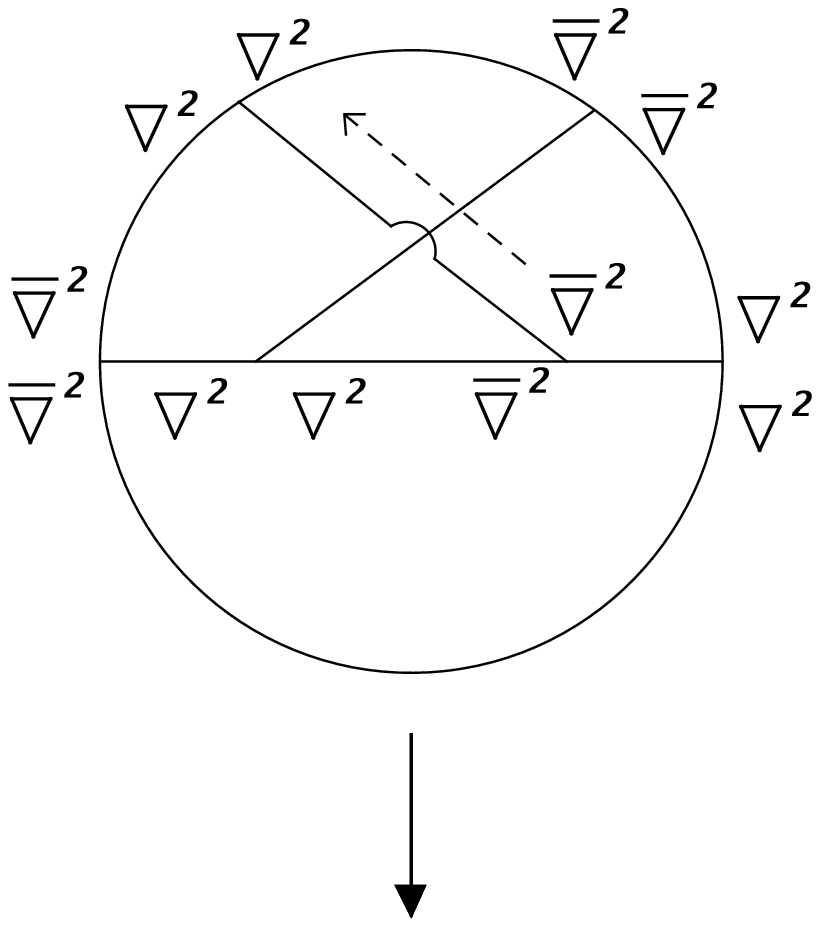} & \\ \\
  \epsfysize=3,3cm\epsfbox{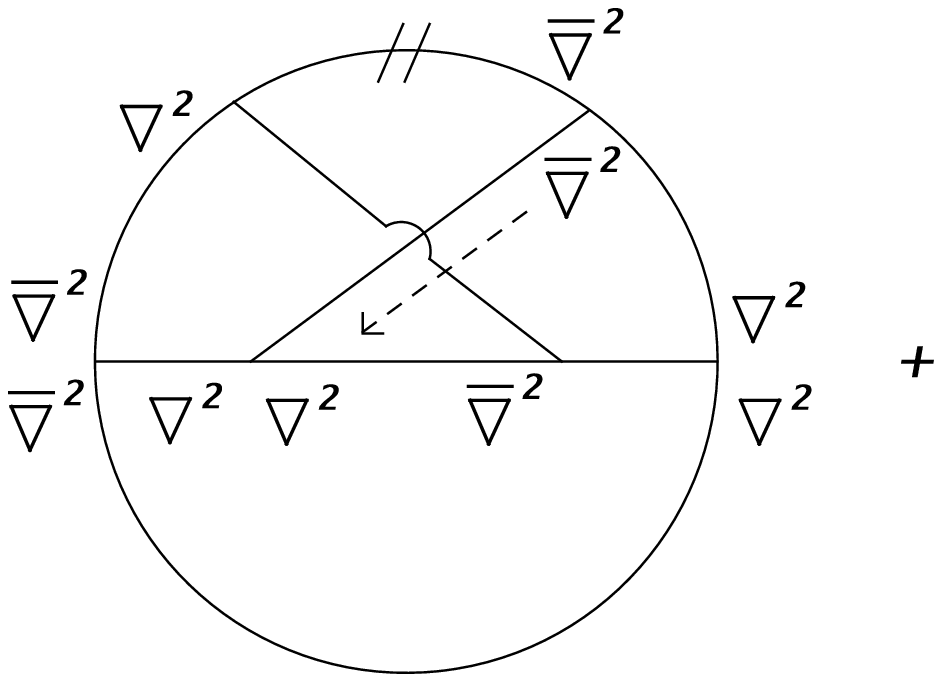} & \epsfysize=3,3cm\epsfbox{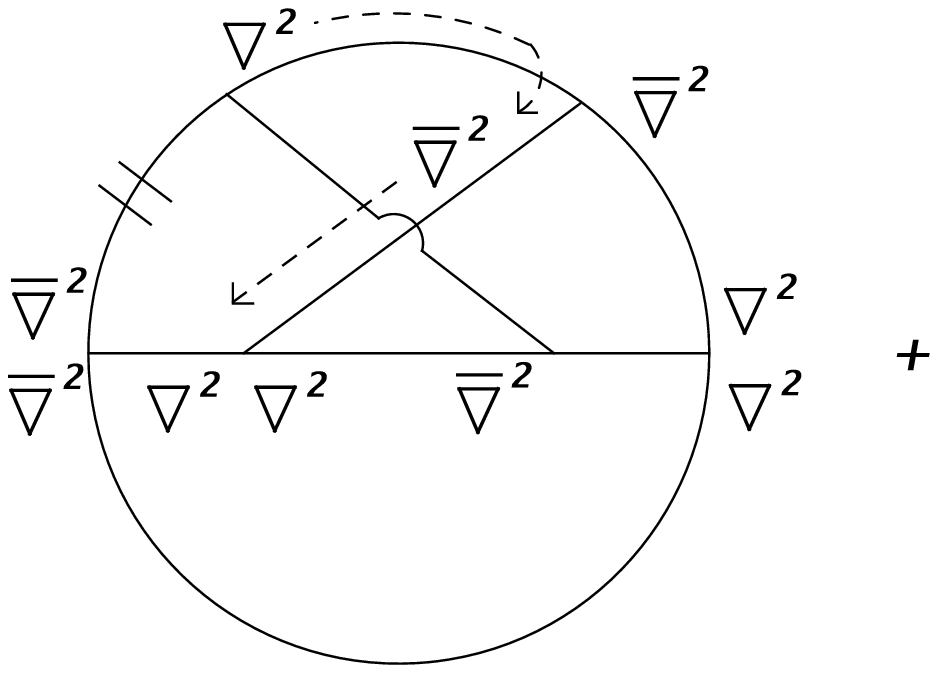}& \epsfysize=3,3cm\epsfbox{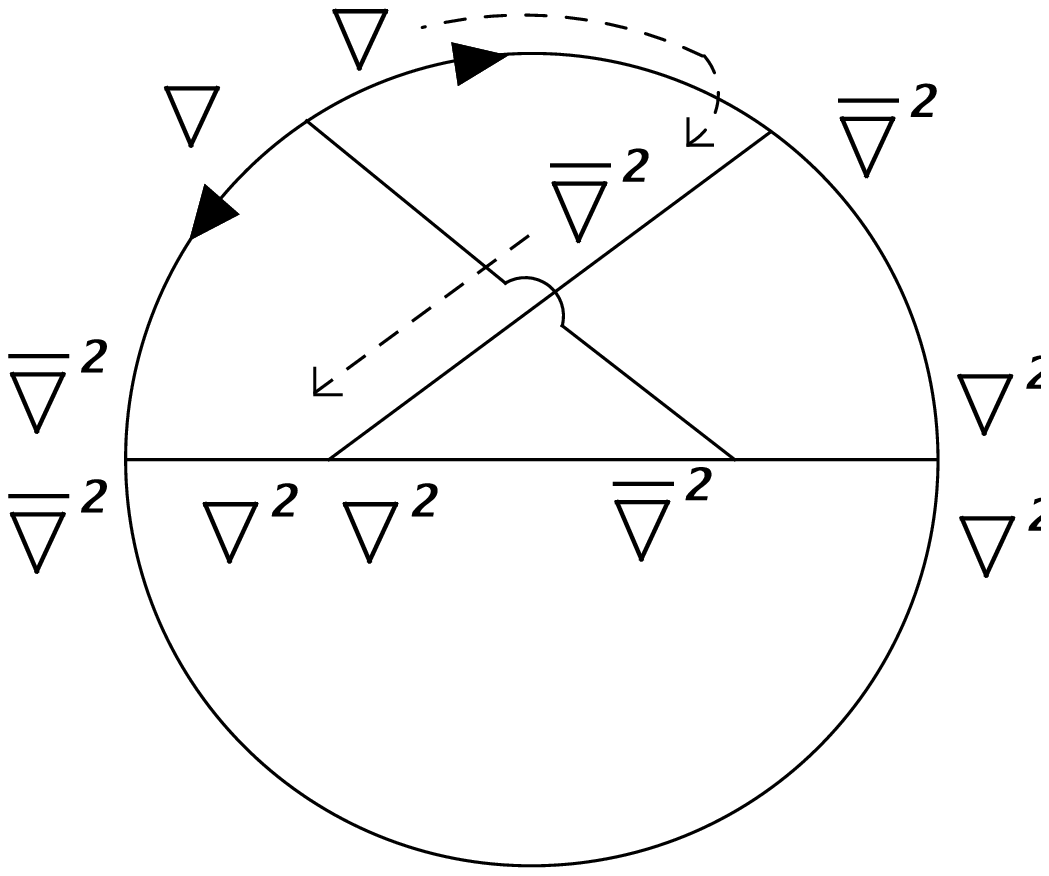} \\
&\hspace{-0,8cm} \epsfysize=1,1cm\epsfbox{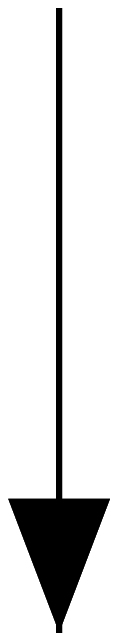} & \\
\epsfysize=3,7cm\epsfbox{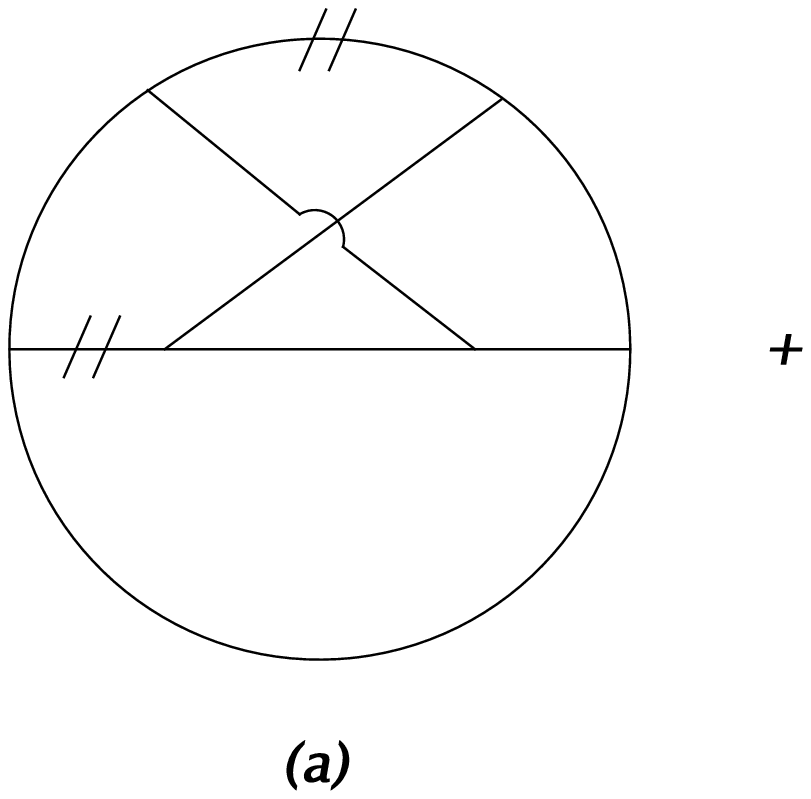} & \epsfysize=3,7cm\epsfbox{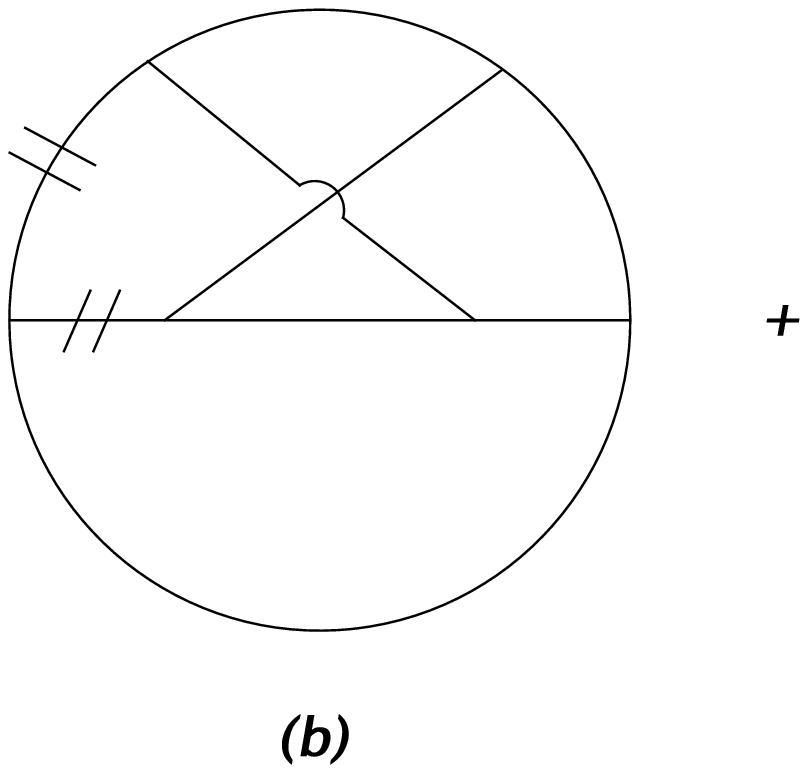}& \epsfysize=3,7cm\epsfbox{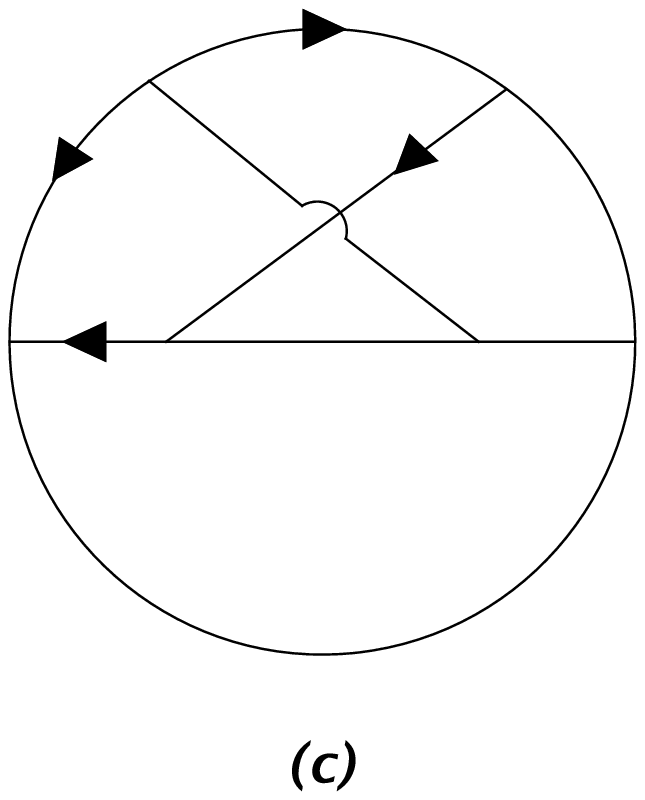}
\end{tabular}
\caption{$\Del$--algebra operations on four--loop vacuum diagram}
\end{center}
\label{vac}
\end{figure}\\
Now we turn to consider the four loop contributions. In this case the
computation is much more involved because we need to perform very non
trivial $\Del$--algebra operations. In \cite{Noi} an analogous problem was
solved by using an alternative procedure, though different from the one
just described which turned
out to be too hard to deal with. Here we want to consider both methods and
show that they indeed give the same result.
 Let us start with the standard procedure. A detailed explanation of
$\Del$--algebra operations can be found in Fig.6. Starting from the top
vacuum diagram and performing integration by parts we end up with three
different graphs. Each of them gives rise to a single bosonic diagram:
Fig.6 (a), (b), (c), where we have denoted
\beq
/\!/ \, \equiv \frac{\frac{1}{2}\Del^a\Del_a}{\Box}~\rightarrow~1~-~\frac{1}{2}~ \frac{\G^a\G_a}{\Box}
\qquad \qquad \Box \equiv \frac{1}{2}\,\partial^a \,\partial_a \qquad \qquad   \blacktriangleright \,\, \equiv  \Del_a=\pa_a-i\G_a
\eeq
 Now we are ready to expand the covariant
propagators to extract tadpole--type contributions.
It is easy to see that (a) and (b) diagrams are equivalent and give rise to
the tadpole graphs shown in Fig.7.
\begin{figure} [ht]
\begin{center}
 \hspace{-1,5cm}\epsfysize=3cm\epsfbox{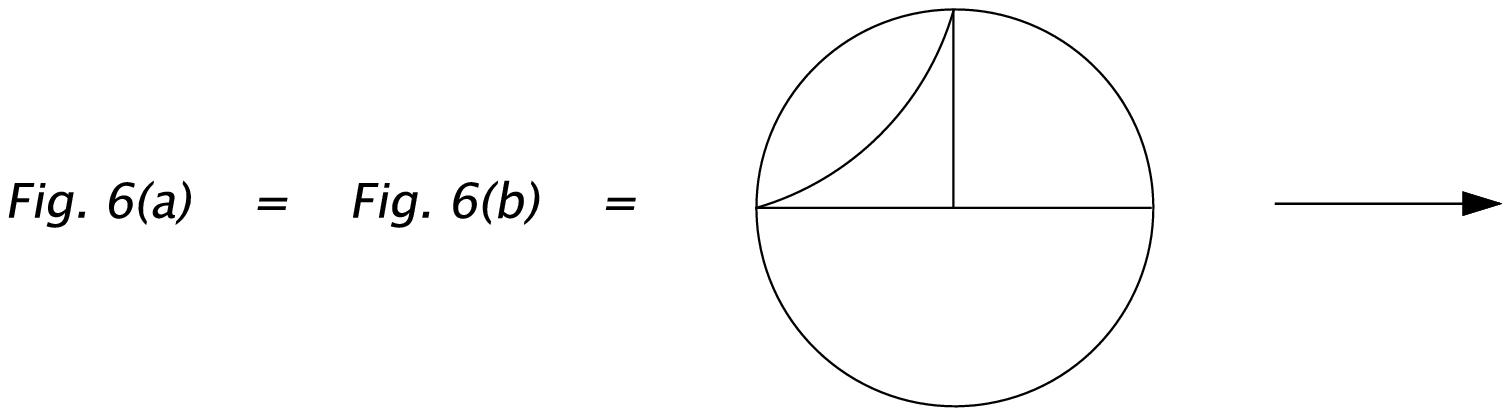}\\
\vspace{0,5cm}
\epsfysize=3,7cm\epsfbox{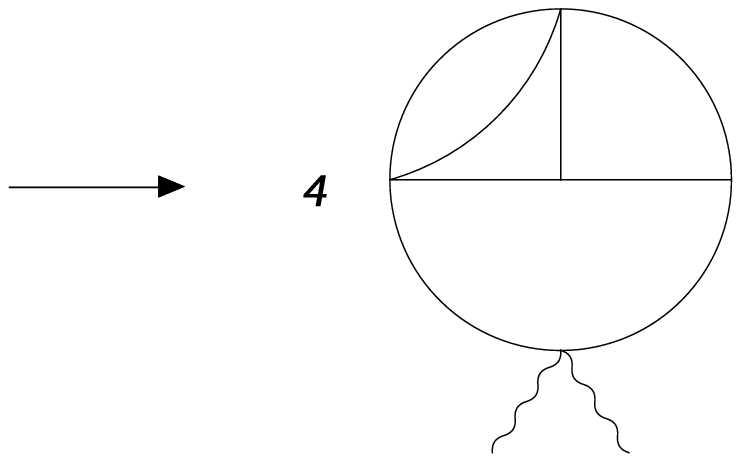} \hspace{0,1cm} \epsfysize=3,7cm\epsfbox{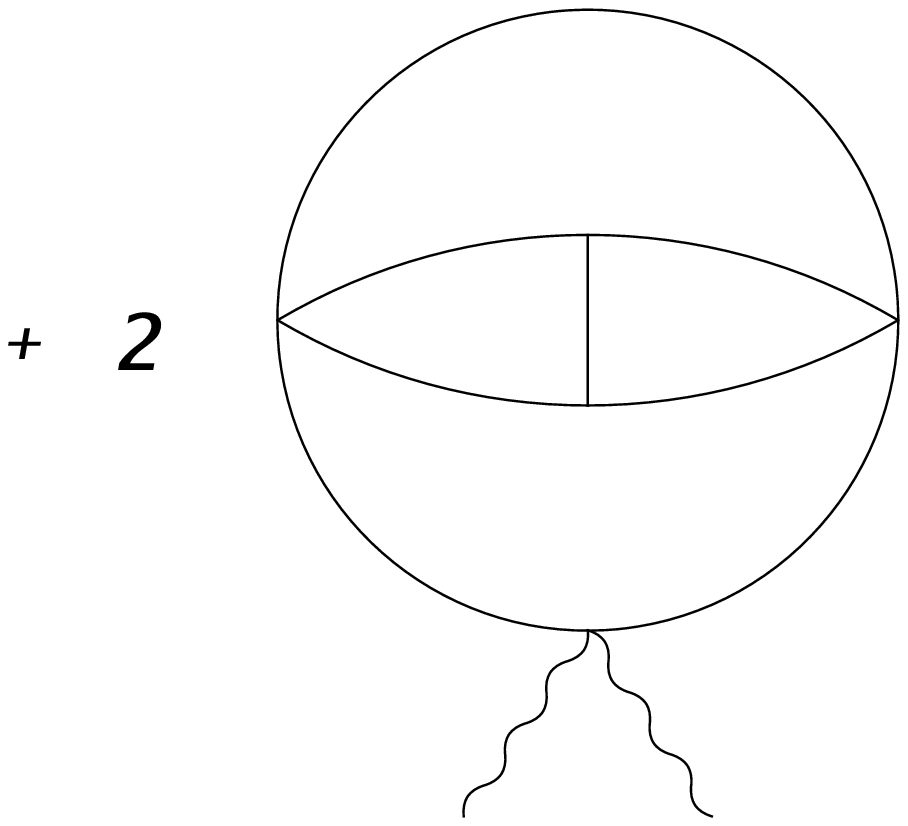} \hspace{0,1cm} \epsfysize=3,7cm\epsfbox{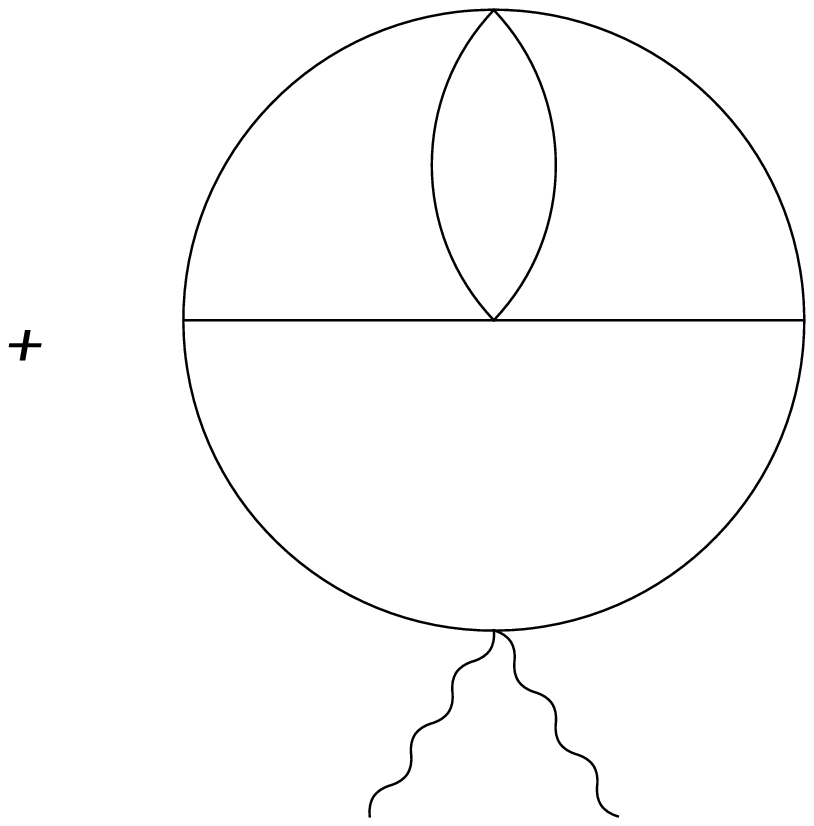}
\vspace{0,3cm}
\caption{Tadpole contributions from propagator expansions of diagrams (a) and (b)}
\end{center}
\label{Adiag}
\end{figure}
\begin{figure} [h]
\begin{center}
 \hspace{-1,1cm}\epsfysize=3cm\epsfbox{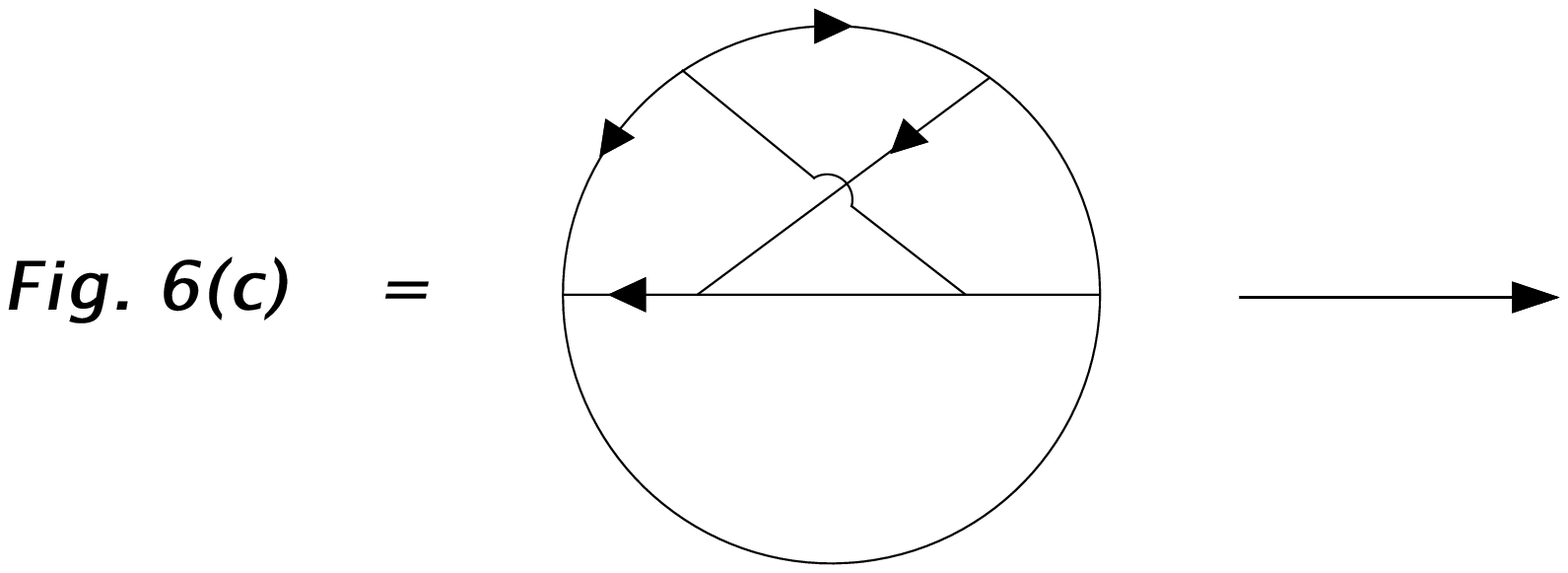} \\
\vspace{0,5cm}
\epsfysize=3,7cm\epsfbox{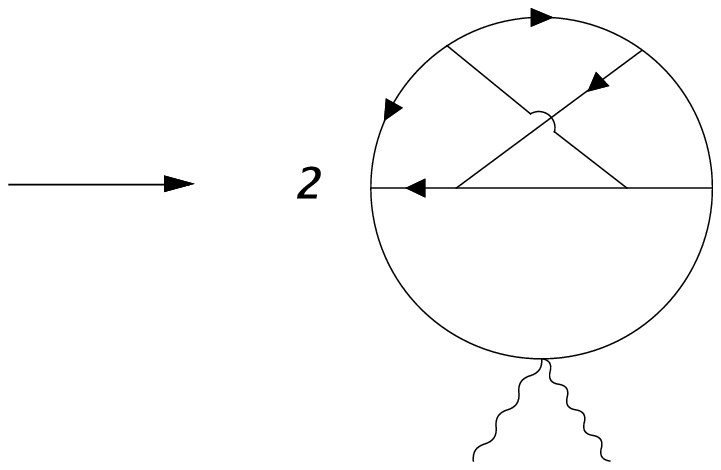} \hspace{0,1cm} \epsfysize=3,7cm\epsfbox{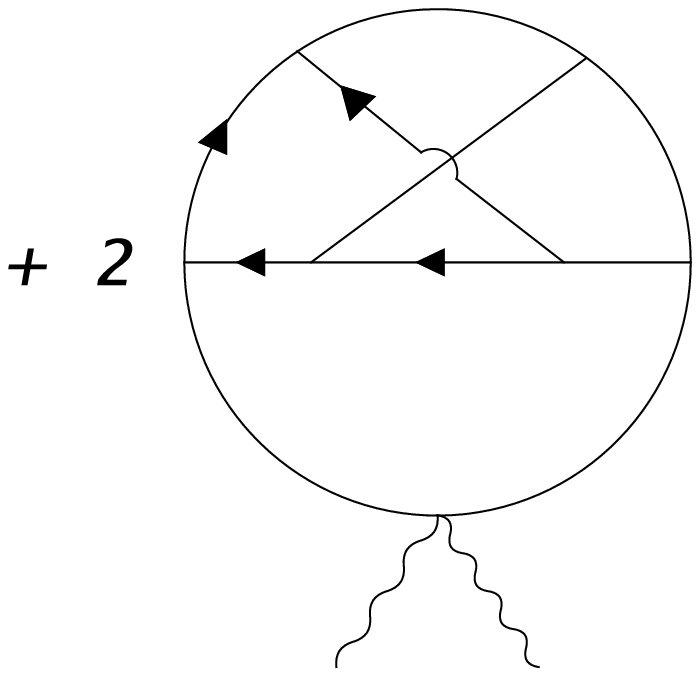}\hspace{0,2cm} \epsfysize=3,7cm\epsfbox{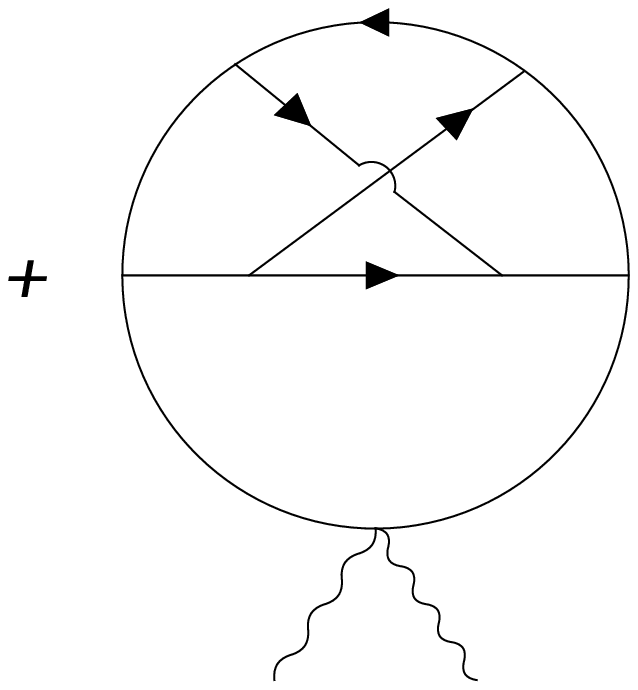}
\vspace{0,3cm}
\caption{Tadpole contributions from relevant propagator expansions of diagram (c)}
\end{center}
\label{Cdiag}
\end{figure}\\
 Analogously the (c) diagram can be expanded to give the
relevant tadpole contributions as indicated in Fig.8.
 The latter integrals are
much  harder to compute because of the presence of four derivatives, indicated
by the black arrows.
 However, after some proper integrations by parts, they can be reduced to
simpler scalar integrals, as depicted in Fig.9. Notice that in the whole procedure we have
neglected  all tadpole graphs with $1/\epsilon$ divergences, which do not
contribute to the four--loop effective action. \\
\begin{figure} [ht]
\begin{center}
\epsfysize=3,7cm\epsfbox{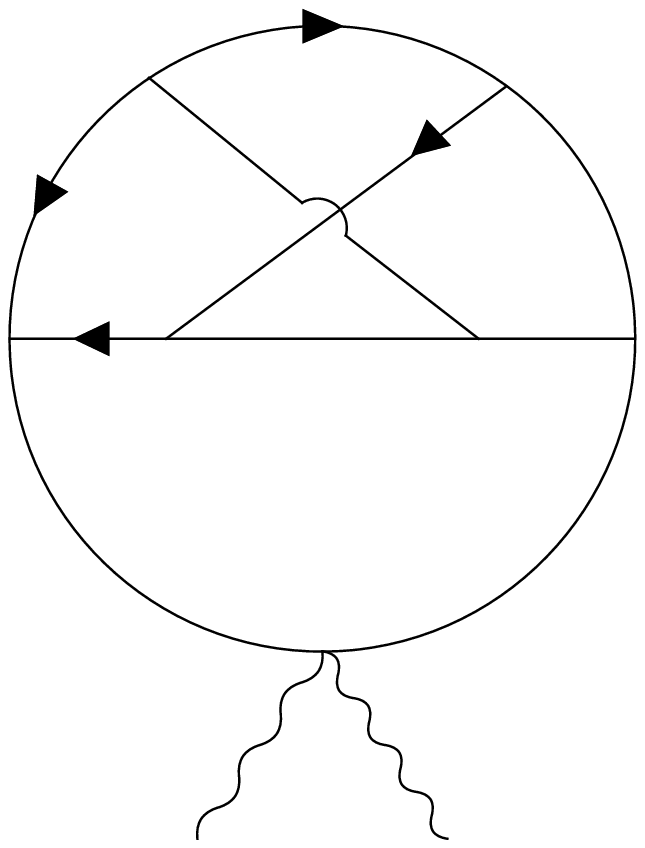} \epsfysize=3,7cm\epsfbox{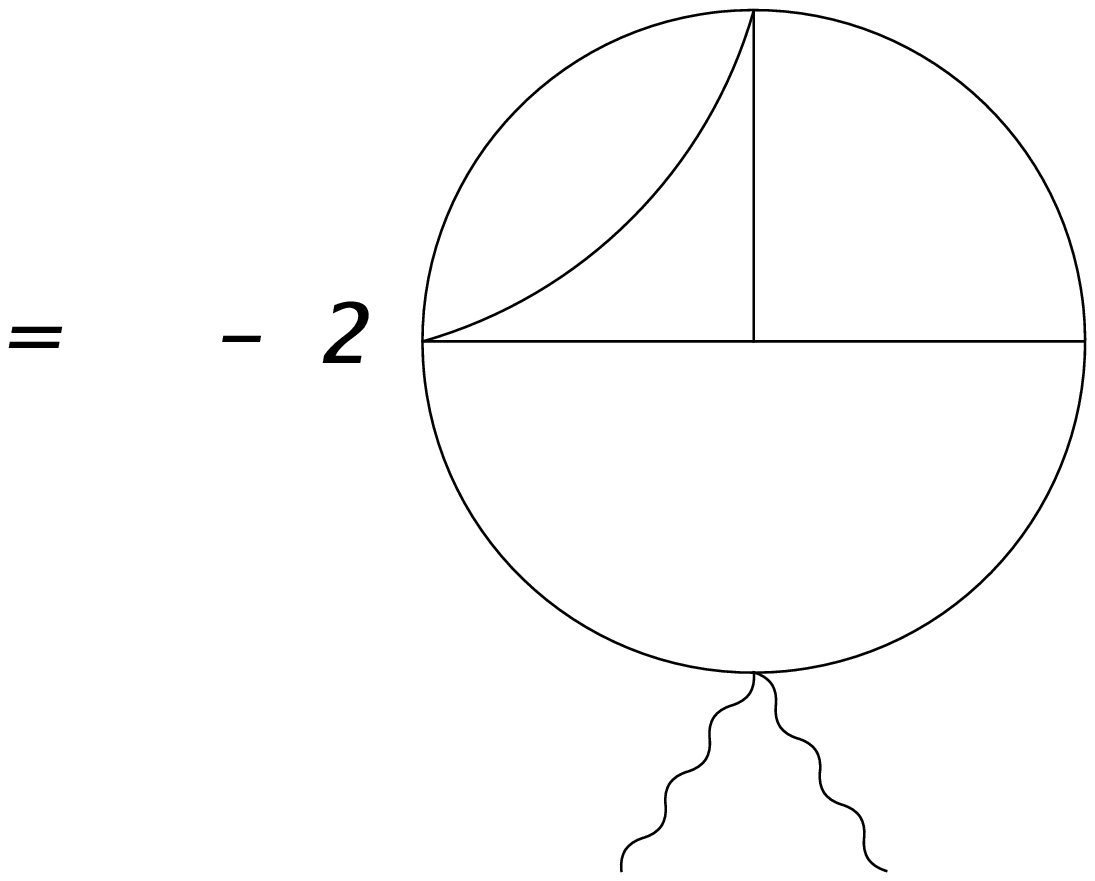} \hspace{0,1cm} \epsfysize=3,7cm\epsfbox{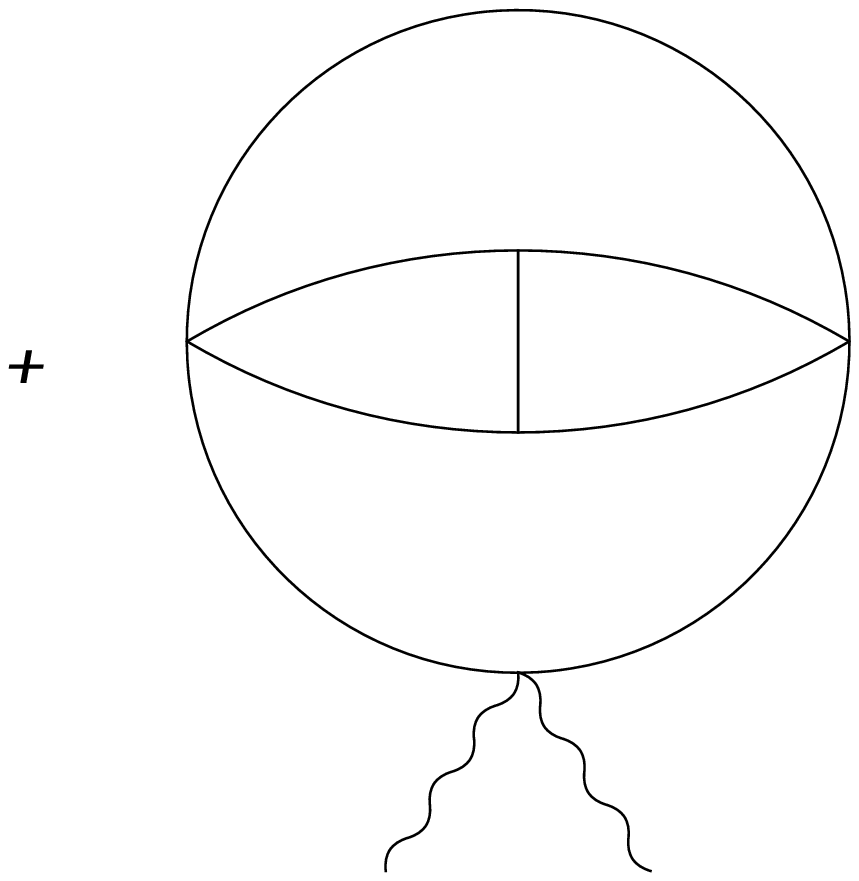} \\
\vspace{0,4cm} \epsfysize=3,7cm\epsfbox{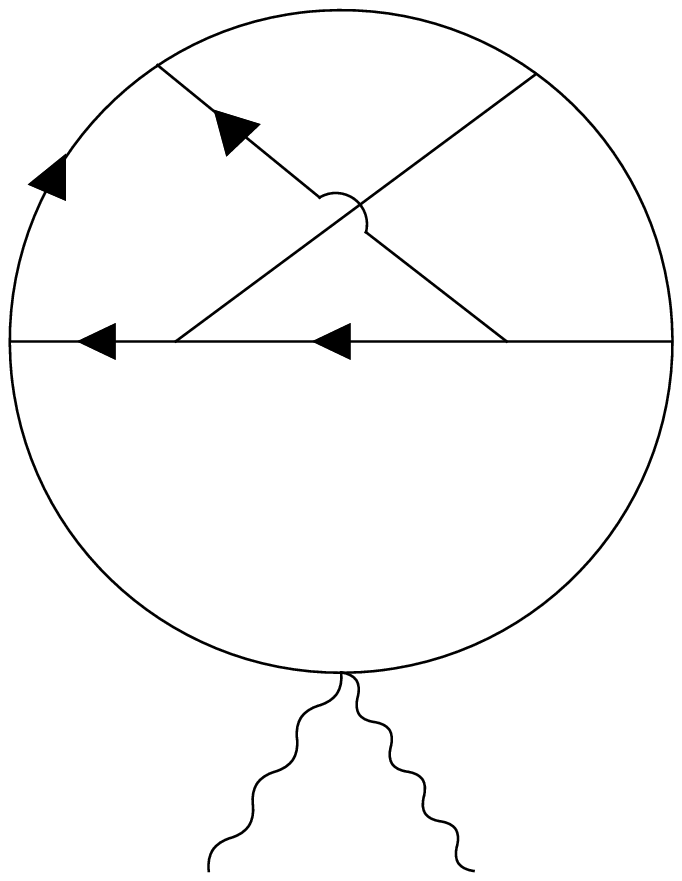} \epsfysize=3,7cm\epsfbox{c1casta.eps} \hspace{0,1cm} \epsfysize=3,7cm\epsfbox{c1Zeta3.eps} \\
\vspace{0,4cm}\hspace{-0,3cm}\epsfysize=3,7cm\epsfbox{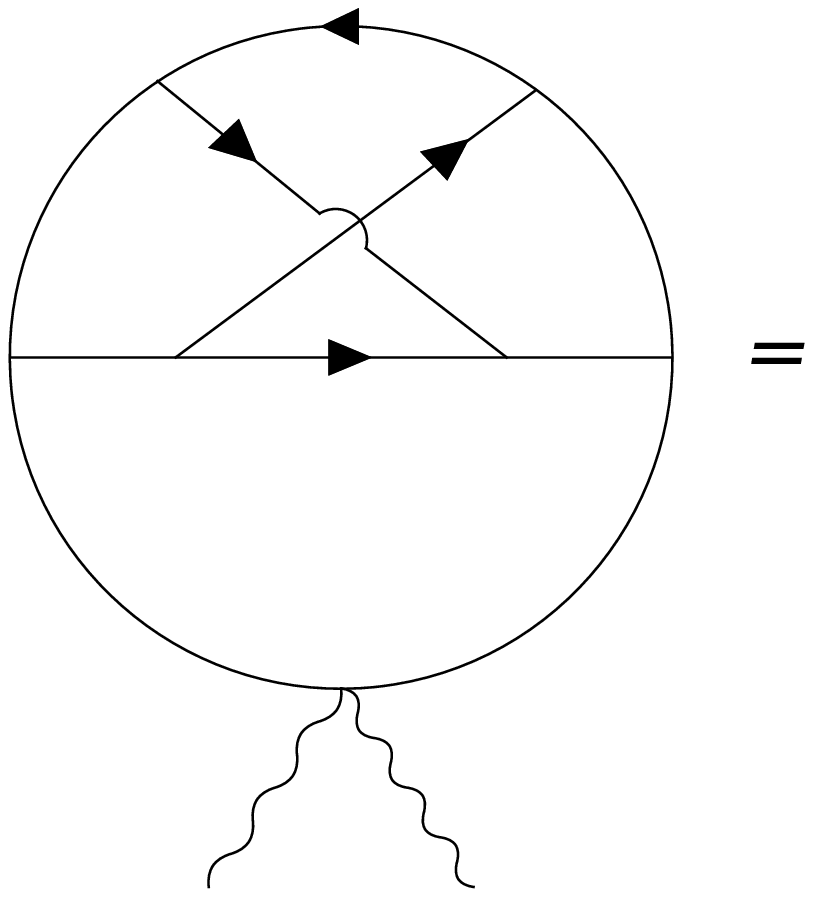}\hspace{0,6cm} \epsfysize=3,7cm\epsfbox{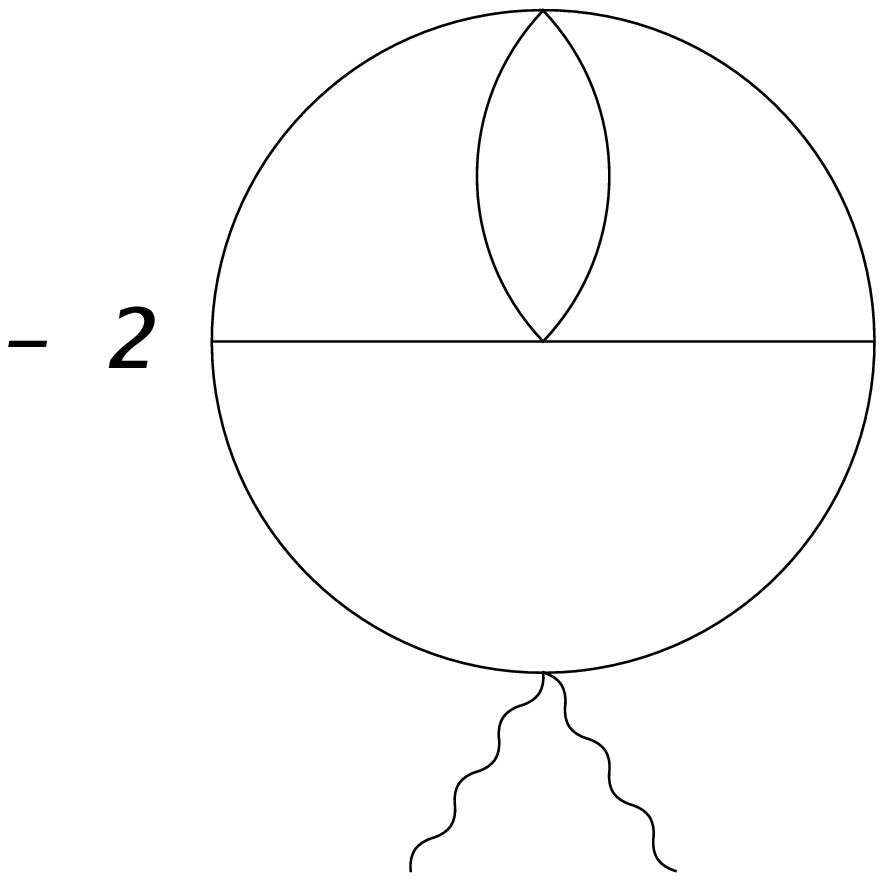} \hspace{0,1cm} \epsfysize=3,7cm\epsfbox{c1Zeta3.eps}
\caption{Scalar reduction of integrals with derivatives}
\end{center}
\label{Cdeve}
\end{figure}
Now we just need to sum up the various contributions generated by (a), (b) and (c) diagrams.
Actually there is no need to compute all these integrals explicitly thanks to a beautiful
diagrammatic cancellation. In fact, the  only surviving terms sum up to give nine times the
same diagram, shown in Fig.10.
 The corresponding bosonic integral is:

\beq
\frac{1}{2}\, \textrm{Tr}(\Gamma^a \Gamma_a)\, \int \frac{d^n  k \,d^n q \,d^n r \,d^n t}{(2 \pi)^{4n}}\, \frac{1}{k^4\, q^2\, t^2\, (r-q)^2 \,(r+t)^2 \,(t+q)^2\, (r+k)^2}
\eeq
\\
So the total four--loop contribution to the effective action, after inserting
color and combinatorial factors and subtracting IR and UV subdivergences  is
given by:

\beq \label{4loop}
\frac{1}{(4 \pi)^2} \,\frac{9 \,(N^2-1)}{8\, N^3}\,B\,\frac{1}{\epsilon}\,\textrm{Tr}
\int d^4 x\, d^2 \theta \,W^\alpha W_\alpha
\eeq
\\
with B defined as in (\ref{B}). This completes the computation of the
four loops contribution with the standard method.\\\\ Had we followed
the alternative procedure developed in \cite{Noi} we would have first
expanded each of the nine propagators of the four--loop vacuum diagram
in Fig.4 and then performed $\Del$--algebra.  In this case, the only
possible contributions would come from two types of diagrams: \\
I. the ones with flat $D^2$ and $\bar{D}^2$ factors at the vertices,
flat propagators and one tadpole insertion, for which now standard
D--algebra can be performed\\ and\\ II. the vacuum diagrams with flat
propagators but $\Del^2$ and $\bar{\Del}^2$ at the chiral vertices in
which the tadpole insertion will have to appear after completion of
the $\Del$--algebra.\\ Analogously to \cite{Noi}, it is easy to see that
only type I diagrams contribute. The computation is now
straightforward. As the vacuum diagram is completely symmetric we have
nine equivalent choices for the propagator to expand. Once a choice
has been made the standard D--algebra gives rise to a unique
contribution, producing precisely the result depicted in Fig.10. We
have therefore checked that as expected the two methods actually give the
same answer.\\
Now we come back to the computation of the gauge beta function and
combine (\ref{2loop}) and (\ref{4loop}). We can easily read the
vanishing condition at order $g^9$:\\

\beq A + \frac{3 \,B}{N^2} = 0 \eeq \\
which is exactly the one
obtained by requiring the vanishing of $\beta_h$ at order $g^7$. Thus we
provide one more confirmation that finiteness theorems for the gauge beta
functions hold even in the CCR context.
 \begin{figure} [ht]
\begin{center}
\epsfysize=4cm\epsfbox{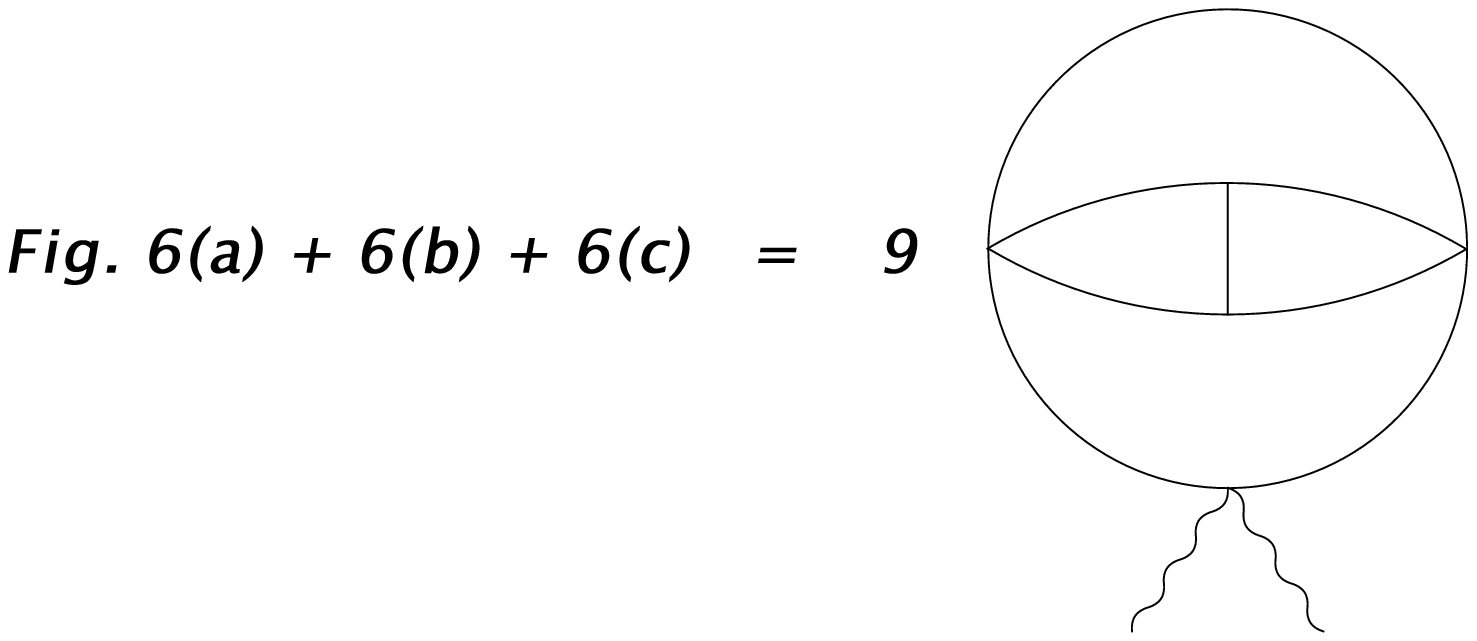}
\caption{Four loop total contribution to the gauge beta function}
\end{center}
\label{Total}
\end{figure}

\section{Conclusions}
In this paper we have considered the $\mathcal{N}=1$ $SU(N)$ super
Yang--Mills theory obtained as a marginal deformation of the
$\mathcal{N}=4$ theory. In particular, we have focused on the
superconformal condition working perturbatively with a complex
deformation parameter $\beta$ at finite $N$.

We have addressed the issue of finding a surface of renormalization
fixed points by requiring the theory to have vanishing beta functions
and using the coupling constant reduction (CCR) procedure. In the CCR
prescription the renormalized chiral couplings are expressed in terms
of a power expansion in the real gauge coupling constant $g$ and this
amounts to face loop mixing at a given order of $g$.

 First, we have concentrated on the chiral beta function ($\beta_h$)
up to $\mathcal{O}(g^7)$. To this end we have fixed the arbitrary
coefficients which appear in the power expansions of the chiral
couplings (\ref{expansion}) by requiring $\gamma=0$ order by order. If
we want to work with a well--defined and a physically meaningful
quantum field theory, we believe that the condition $\beta_h=0$ should
not be affected by scheme dependence. Scheme independence of the
conformal definition of the theory introduces a further constraint on
the couplings. Here comes the novelty with respect to the planar case
studied in \cite{Noi}. The planar limit involves only two of the three
independent constants in (\ref{h123}) and scheme independence of the
theory forces $\beta$ to be {\it real}. On the other hand, keeping $N$
finite, all of the three parameters $\hu,\hd,\htr$ enter the
superconformal condition allowing for a {\it complex} deformed theory
which is scheme--independent at least at $\mathcal{O}(g^{10})$. We expect
this pattern should hold even for higher orders.

Then we have considered the gauge beta function $\beta_g$. Working in
the CCR context we are not guaranteed that standard finiteness
theorems \cite{PW,GMZ} are valid. In \cite{Noi} a generalization of
these theorems was proposed: if $\beta_h=0$ up to
$\mathcal{O}(g^{2n+1})$ then $\beta_g=0$ up to
$\mathcal{O}(g^{2n+3})$. This statement was checked in the planar
limit for $n=4$ using an alternative procedure for covariant
$\nabla$--algebra. Here we have provided another highly non--trivial
confirmation of this proposal in the non--planar theory for
$n=3$. Moreover, we have explicitly checked that the simplified
$\nabla$--algebra technique used in \cite{Noi} is equivalent to the
standard one.
\newpage

\section*{Appendix A}

In this Appendix we report the full expression for the color of the four loop diagram depicted in Fig.3:
\bea
 K_4 & = &\frac{1}{2}\Big[(|h_1|^2+|h_2|^2)^4+(|h_1|^2-|h_2|^2)^4\Big] + \non \\ \non \\
& + & \frac{4}{N^2}\Big[|h_3|^8-4|h_3|^6(|h_1|^2+|h_2|^2)+ 2|h_3|^4(3|h_1|^4+ 4|h_1|^2|h_2|^2+3|h_2|^4)+ \non \\\non \\
& -& 2 |h_3|^2 (3 |h_1|^6 + 5 |h_1|^4|h_2|^2 + 5 |h_1|^2 |h_2|^4 + 3 |h_2|^6)+ \non \\\non \\
& + & (|h_1|^8 + 8 |h_1|^6|h_2|^2 + 6 |h_1|^4 |h_2|^4 + 8 |h_1|^2 |h_2|^6 + |h_2|^8)\Big]+ \non \\ \non \\
& -& \frac{4}{N^4}\Big[5|h_3|^8-20|h_3|^6(|h_1|^2+|h_2|^2)+12 |h_3|^4(|h_1|^4+|h_1|^2|h_2|^2+|h_2|^4)+\non \\ \non \\
&- & 8 |h_3|^2 (|h_1|^6 - |h_1|^4|h_2|^2 - |h_1|^2 |h_2|^4 + |h_2|^6) \Big] +\non  \\\non \\
&-& \frac{4}{N^6}\Big[10|h_3|^8+32|h_3|^6(|h_1|^2+|h_2|^2)-8|h_3|^4 |h_1|^2 |h_2|^2
\Big] + \frac{256}{N^8} |h_3|^8 \non
\eea
From this formula one can easily obtain the explicit value of the $f$ function  in (\ref{uxi}):

\bea
f & = & 8\Big[a_1^4 + 8 a_1^3 b_1 + 6 a_1^2 b_1^2 + 8 a_1 b_1^3 + b_1^4 - 2(a_1 + b_1)(3 a_1^2 + 2 a_1 b_1 + 3 b_1^2)c_1 + \non \\
& +& 2(3 a_1^2 + 4 a_1 b_1 + 3 b_1^2) c_1^2 - 4(a_1 + b_1) c_1^3 + c_1^4\Big] + \frac{8}{N^2} \Big[8(a_1 - b_1)^2
(a_1 + b_1) c_1 + \non \\
& -& 12 (a_1^2 + a_1 b_1 + b_1^2) c_1^2 + 20(a_1 + b_1) c_1^3 - 5 c_1^4\Big] + \frac{8}{N^4}\Big[8 a_1 b_1 c_1^2
- 32(a_1 + b_1) c_1^3 - 10 c_1^4\Big] + \non \\
& +& \frac{512}{N^6} c_1^4 \non
\eea

\vspace{0,5cm}
\noindent
{\bf Acknowledgments}\\
We are grateful to S.~Penati, A.~Santambrogio, D.~Zanon and G.~Tartaglino--Mazzucchelli for useful discussions and important comments.\\
This work has been supported in part by INFN, PRIN prot. 2005024045-002 and the European Commission RTN program MRTN-CT-2004-005104.

\end{document}